\documentclass[11pt,showpacs,superscriptaddress,amsmath,amssymb,nofootinbib]{revtex4}
\usepackage{graphicx}
\usepackage{dcolumn}
\usepackage{bm}
\usepackage{amssymb}
\usepackage{amsmath}
\usepackage{epsfig}
\usepackage{color}
\usepackage{slashed}
\usepackage{float}
\usepackage{longtable}
\usepackage{braket}
\usepackage{ulem}

\begin{document}

\title{Can the 750-GeV diphoton resonance be the singlet Higgs boson of \\
custodial Higgs triplet model?}

\author{Cheng-Wei Chiang}
\email{chengwei@ncu.edu.tw}
\affiliation{Institute of Physics, Academia Sinica, Taipei, Taiwan 11529, Republic of China}
\affiliation{Department of Physics and Center for Mathematics and Theoretical Physics,
National Central University, Chungli, Taiwan 32001, Republic of China}
\affiliation{Physics Division, National Center for Theoretical Sciences, Hsinchu, Taiwan 30013, Republic of China}
\author{An-Li Kuo}
\email{101222028@cc.ncu.edu.tw}
\affiliation{Department of Physics and Center for Mathematics and Theoretical Physics,
National Central University, Chungli, Taiwan 32001, Republic of China}

\begin{abstract}
The observation of diphoton excess around the mass of 750~GeV in LHC Run-II motivates us to consider whether the singlet Higgs boson in the custodial Higgs triplet model can serve as a good candidate because an earlier study of comprehensive parameter scan shows that it can have the right mass in the viable mass spectra.  By assuming the singlet Higgs mass at 750~GeV, its total width less than 50~GeV and imposing constraints from the LHC 8-TeV data, we identify an approximately linear region on the $(v_\Delta, \alpha)$ plane along which the exotic Higgs boson masses satisfy a specific hierarchy and have lower possible spectra, where $v_\Delta$ denotes the triplet vacuum expectation value and $\alpha$ is the mixing angle between the singlet Higgs boson and the standard model-like Higgs boson.  
Although the diphoton decay rate can be enhanced by charged Higgs bosons running in the loop in this region, it is mostly orders of magnitude smaller than that required for the observed production rate, except for the small $v_\Delta$ region when the diphoton fusion production mechanism becomes dominant.  Nonetheless, this part of parameter space suffers from the problems of breakdown of perturbativity and large uncertainties in the photon parton distribution function of proton.
\end{abstract}


\maketitle
\newpage

\section{Introduction}

Latest data of high-mass diphoton events from both ATLAS and CMS in LHC Run-II point us to an tantalizing possibility of having a resonance around the mass of 750~GeV~\cite{ATLAS,CMS:2015dxe}.  Collected using  $3.2$~fb$^{-1}$ and   $2.6$~fb$^{-1}$ of data, respectively, the two collaborations observe signals with the significances of $3.9\sigma$ and $3.4\sigma$.  Assuming a narrow total width, the measurements are consistent with the cross sections of
\begin{align}
\sigma(pp\to S \to \gamma\gamma) =
\left\{
\begin{array}{l}
(5.5 \pm 1.5)~{\rm fb}~~ ({\rm ATLAS}) ~,
\\
(4.8 \pm 2.1)~{\rm fb}~~ ({\rm CMS}) ~,
\end{array}
\right.
\label{eq:data}
\end{align}
respectively~\cite{Buttazzo:2015txu}, where $S$ denotes the putative 750-GeV resonance.

Among the many proposals for this resonance, quite a few consider the scenario that $S$ is a singlet scalar boson that mixes with the 125-GeV Higgs boson~\cite{Buttazzo:2015txu,Mambrini:2015wyu}.  As an interesting alternative, such a singlet scalar can have its origin from the custodial Higgs triplet model (also known as the Georgi-Machacek model)~\cite{Georgi:1985nv,Chanowitz:1985ug}.  In addition to the standard model (SM) Higgs doublet, the custodial Higgs triplet model also introduces a complex and a real Higgs triplet fields.  By imposing a global $SU(2)_L \times SU(2)_R$ symmetry on the Higgs potential and with an alignment in the vacuum expectation value (VEV) of both triplet fields, the electroweak $\rho$ parameter can be kept at unity at tree level even when the triplet VEV is sizeable.  The model is well-motivated because the triplet VEV can give rise to Majorana mass for neutrinos, the so-called type-II seesaw mechanism.  After electroweak symmetry breaking, the triplet fields are decomposed into a quintet, a triplet, and a singlet states under the SM SU(2)$_L$ group~\cite{Gunion:1989ci}.  Although the triplet fields are introduced to couple with only the left-handed leptons among the SM fermions, the triplet and singlet states can couple with other SM particles through respective mixing with the Goldstone and Higgs bosons from the SM Higgs doublet field.  In particular, the heavier Higgs singlet boson, denoted by $H_1^0$, may be a candidate for the 750-GeV resonance.

In a recent analysis~\cite{Chiang:2015amq}, we had performed a comprehensive scan of the parameters in the custodial Higgs triplet model and obtained the parameter space that was consistent with the measured signal strengths of the 125-GeV Higgs boson, oblique $S$ parameter, $Zb \bar b$ coupling, as well as the perturbative unitarity and vacuum stability bounds.  From the viable Higgs mass spectra, we found that indeed $H_1^0$ could have a mass of 750~GeV with an appropriate choice of parameters and still be consistent with the above-mentioned constraints.  Based on the previous findings, we discuss in this work whether $H_1^0$ also has the capacity to explain the observed diphoton excess and the inferred total width.  After imposing the constraints of 8-TeV searches in various channels, we point out that the masses of the physical quintet $H_5$'s and triplet $H_3$'s as well as the triplet VEV $v_\Delta$ and the singlet mixing angle $\alpha$ in this case are restricted to specific ranges.  In particular, the range of $v_\Delta$ is consistent with other constraints~\cite{Chiang:2014bia,Chiang:2015amq}.  We note that though with some dependence on $v_\Delta$ and $\alpha$, the masses of $H_5$'s and $H_3$'s have a specific correlation.  For a fixed $v_\Delta$ ($\alpha$), increasing $\alpha$ $(v_\Delta)$ will make the exotic Higgs boson masses tends to follow the hierarchy $m_{H_1} > m_{H_3} > m_{H_5}$, which can affect our strategy of searching for the other exotic Higgs bosons at the LHC~\cite{Chiang:2012cn}. With the lowest mass bounds of $m_{H_5}\sim280$~GeV and $m_{H_3}\sim650$~GeV, we are able to study the decays of $H_1^0$ with more certainty.  Using such information and based upon the physically allowed parameter space given in Ref.~\cite{Chiang:2015amq}, we find that in most of the parameter space the singlet Higgs boson cannot explain the diphoton excess without invoking additional particles or mechanisms to enhance the production or diphoton decay.  However, for a small region of $(\alpha,v_\Delta)$ close to the origin, the partial width of $H_1^0 \to \gamma\gamma$ is significantly enhanced and the $\gamma\gamma$ fusion production mechanism becomes dominant.  There is then a possibility to explain the diphoton signal, as also noted in Ref.~\cite{Fabbrichesi:2016alj}.  In such cases, however, the one-loop correction to the quartic $H_1$ coupling mediated by the $H_5$ bosons becomes too large to grant the validity of perturbative calculations.  Moreover, the photon parton distribution function in the proton suffers from large uncertainties, particularly in the large momentum fraction region ($\sim 30\%$ to $60\%$ for $x \sim 0.001$ to $0.1$), so that the prediction of $\gamma\gamma$ fusion contribution becomes unreliable.

This paper is organized as follows.  In Section~\ref{sec:review}, we briefly review the custodial Higgs triplet model and the constraints analyzed in Ref.~\cite{Chiang:2015amq}.  In Section~\ref{sec:excess}, we use data of various search channels to further constrain the parameter space, particularly the mass spectrum.  We show that within the allowed space, the maximal diphoton production through $H_1^0$ is still off the observed diphoton excess by 1 to 3 orders of magnitude when only the digluon fusion and vector boson fusion processes are taken into account.  When the diphoton fusion process is also included, the diphoton signal can be enhanced in a small region of $(\alpha,v_\Delta)$ close to the origin.  We then show that this region suffers from the problems of breakdown of perturbativity and large uncertainties in the photon parton distribution function of proton.  Section~\ref{sec:summary} summarizes our findings.

\section{Review of the custodial Higgs triplet model and major constraints \label{sec:review}}

The custodial Higgs triplet model extends the SM Higgs sector with two weak isospin triplet scalar fields with hypercharge $Y=1$ and $Y=0$~\footnote{The relation between the electric charge $Q$, the third component of the weak isospin $I_3$ and the hypercharge $Y$ is $Q = I_3 + Y$.}.  Writing in the $SU(2)_L \times SU(2)_R$-covariant form, the doublet and triplet fields are respectively
\begin{align}
\Phi=\left(
\begin{array}{cc}
\phi^{0*} & \phi^+ \\
-\phi^- & \phi^0
\end{array}\right) ~,~~
\Delta=\left(
\begin{array}{ccc}
\chi^{0*} & \xi^+ & \chi^{++} \\
-\chi^- & \xi^0 & \chi^{+} \\
\chi^{--} & -\xi^- & \chi^{0} 
\end{array}\right) ~.
\label{eq:Higgs_matrices}
\end{align}

The most general Higgs potential invariant under the global $SU(2)_L\times SU(2)_R\times U(1)_Y$ symmetry is
\begin{align}
\begin{split}
V
=& \ \frac{1}{2} m_1^2 \, {\rm tr}[ \Phi^{\dagger} \Phi ] + 
\frac{1}{2} m_2^2 \, {\rm tr}[ \Delta^{\dagger} \Delta ]
 +  \lambda_1 \left( {\rm tr}[ \Phi^{\dagger} \Phi ] \right)^2 
 +  \lambda_2 \left( {\rm tr}[ \Delta^{\dagger} \Delta ] \right)^2 
\\
&
 +  \lambda_3 {\rm tr}\left[ \left( \Delta^{\dagger} \Delta \right)^2 \right] 
 +  \lambda_4 {\rm tr}[ \Phi^{\dagger} \Phi ] {\rm tr}[ \Delta^{\dagger} \Delta ]
 +  \lambda_5 {\rm tr}\left[ \Phi^{\dagger} \frac{\sigma^a}{2} \Phi \frac{\sigma^b}{2} \right] 
                  {\rm tr}[ \Delta^{\dagger} T^a \Delta T^b]
\\
 &+ \mu_1 {\rm tr}\left[ \Phi^{\dagger} \frac{\sigma^a}{2} \Phi \frac{\sigma^b}{2} \right]
                               (P^{\dagger} \Delta P)_{ab}
 + \mu_2 {\rm tr}[ \Delta^{\dagger} T^a \Delta T^b]
                               (P^{\dagger} \Delta P)_{ab} ~,
\end{split}
\label{potential}
\end{align}
where summations over $a,b = 1,2,3$ are understood, $\sigma$'s and $T$'s are respectively the Pauli matrices and the $3\times3$ matrix representation of the SU(2) generators, and
\begin{align}
P &= \frac{1}{\sqrt{2}} \left( \begin{array}{ccc}
-1 & i & 0 \\
0 & 0 & \sqrt{2} \\
1 & i & 0
\end{array}
\right)  \nonumber
\end{align}
diagonalizes the adjoint representation of the SU(2) generators.  It is noted that all parameters in the Higgs potential are real and thus do not allow CP violation.

The triplet fields attain VEV's induced by the breakdown of electroweak symmetry.  With vacuum alignment in the triplet fields, $\langle \chi^0 \rangle = \langle \xi^0 \rangle \equiv v_\Delta$, the model preserves the custodial symmetry and keeps the electroweak $\rho$ parameter at unity at tree level.  The doublet and triplet VEV's satisfy the relation $v_{\Phi}^2+8v_{\Delta}^2 = (246~{\rm GeV})^2$.

The global $SU(2)_L\times SU(2)_R$ symmetry of the Higgs potential is explicitly broken by the Yukawa and the hypercharge interactions to the custodial $SU(2)_L$ symmetry, under which the $\Delta$ field is decomposed into the $\bf 5$, $\bf 3$, and $\bf 1$ representations while the $\Phi$ field is decomposed into the $\bf 3$ and $\bf 1$ representations.  The $\bf 5$ representation is CP-even.  Through the mixing angle $\beta$ defined by $\tan \beta \equiv v_{\Phi}/ \left( 2\sqrt{2}v_{\Delta} \right)$, the two $\bf 3$ representations mix with each other to give a physical CP-odd 3-plet $H_3$ and Goldstone bosons that become the longitudinal components of the $W$ and $Z$ bosons.  The two CP-even $\bf 1$ representations, on the other hand, mix with each other to give the 125-GeV SM-like Higgs boson $h$ and another singlet $H_1^0$ through another mixing angle $\alpha$, the explicit definition of which can be found in Ref.~\cite{Chiang:2015amq}.  Due to the custodial symmetry, particles within the same multiplet have the same mass, neglecting ${\cal O}(100)$~MeV mass difference among different charged states.  We therefore collectively denote the masses of physical 5-plet, 3-plet, and singlet exotic Higgs bosons by $m_{H_5}$, $m_{H_3}$, and $m_{H_1}$, respectively.

\begin{table}
\centering
\begin{tabular}{ccc}
\hline\hline
Higgs & ~~~~$\kappa_F$~~~~ 
& ~~~~~~~~~~~$\kappa_V$~~~~~~~~~~~ \\
\hline
$h$ & $\displaystyle \frac{\cos\alpha}{\sin\beta}$
& $\displaystyle \sin\beta\cos\alpha - \sqrt{\frac83} \cos\beta\sin\alpha$
\\
$H_1^0$ & $\displaystyle \frac{\sin\alpha}{\sin\beta}$ 
& $\displaystyle \sin\beta\sin\alpha + \sqrt{\frac83} \cos\beta\cos\alpha$
\\
$H_3^0$ & $i\eta_f\cot\beta$ 
& 0
\\
$H_5^0$ & 0 
& $\displaystyle \kappa_W = -\frac{\cos\beta}{\sqrt{3}}$ and 
$\displaystyle \kappa_Z = \frac{2\cos\beta}{\sqrt{3}}$
\\
\hline\hline
\end{tabular}
\\
\caption{Scaling factors for the couplings between the neutral Higgs bosons in the custodial Higgs triplet model and SM fermions and weak gauge bosons, as compared to the corresponding couplings of the SM Higgs boson.  $\eta_f = +1$ for up-type quarks and $-1$ for down-type quarks and charged leptons.
\label{coupling}}
\end{table}

Table~\ref{coupling} summarizes how the neutral Higgs bosons in the model couple to the SM fermions and weak gauge bosons, as compared to the corresponding couplings of the SM Higgs boson.  They are expressed in terms of the scaling factors defined as the ratios of Higgs couplings:
\begin{align}
\kappa_F[\phi] = \frac{g_{\phi f \bar f}}{g_{h_{SM} f \bar f}}
~,~~
\kappa_V[\phi] = \frac{g_{\phi VV}}{g_{h_{SM} VV}}
~,
\end{align}
where $\phi = h$, $H_1^0$, $H_3^0$ or $H_5^0$, and $h_{SM}$ denotes the SM Higgs boson.  It is noted that $H_3^0$ is gauge-phobic while $H_5^0$ is fermio-phobic.

\begin{figure}[ht]
\centering
\includegraphics[scale=0.7]{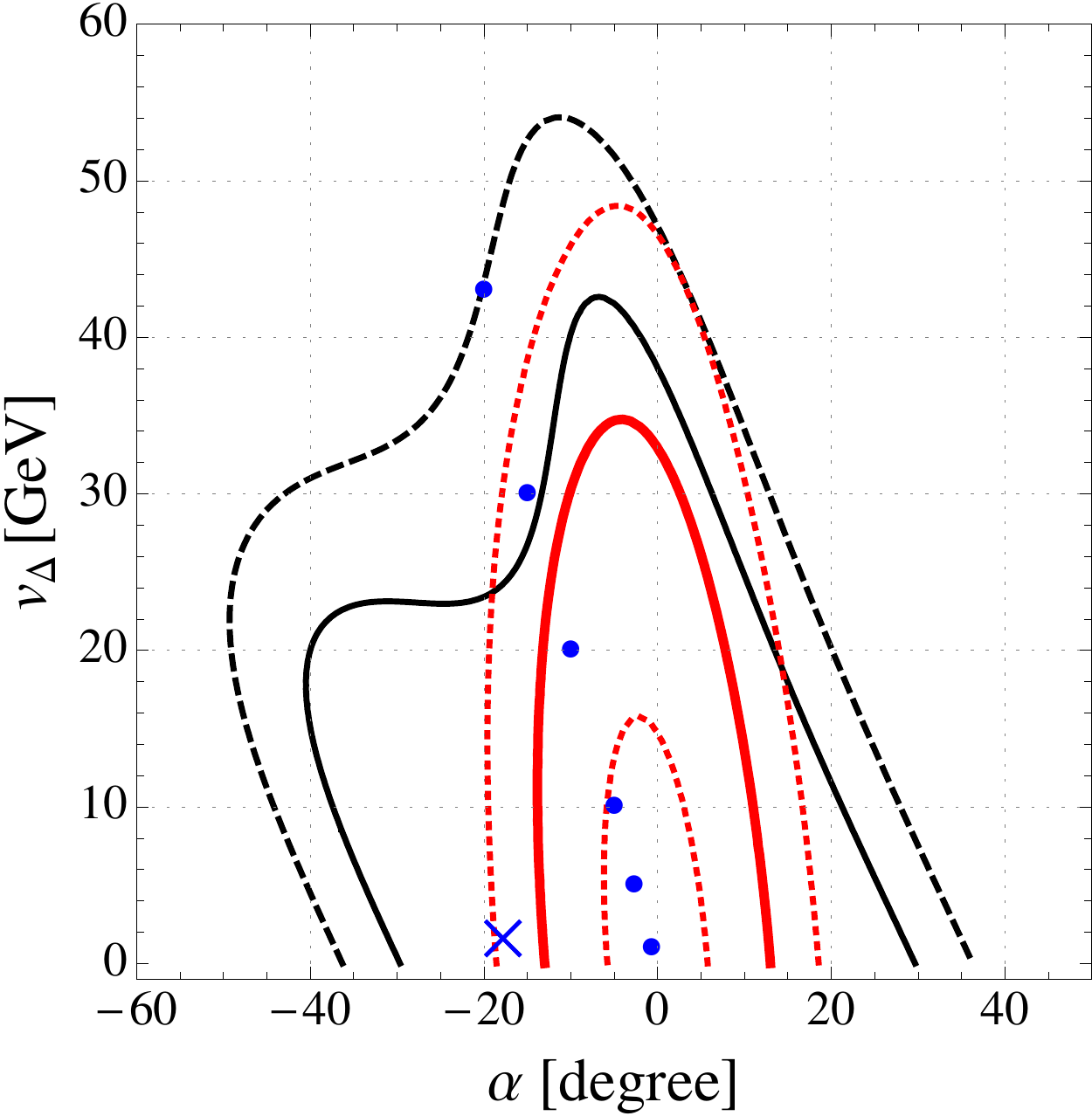}
\caption{The thick black solid and dashed curves are respectively the $1\sigma$ and $2\sigma$ contours of a $\chi^2$ fit to the SM-like Higgs boson signal strengths, except for the $\gamma\gamma$ channel.  The $\chi^2$ minimum is marked by the blue cross.  From inside out, the red contours correspond to the partial production cross section $\sigma(pp\to H_1^0)_{\rm GGF+VBF}=10$, 50 and 100~fb at the 13-TeV LHC, where the K-factors have been taken into account.  The blue benchmark points used in this analysis lie roughly along a line.}
\label{chipptoH1}
\end{figure}

Ref.~\cite{Chiang:2015amq} scanned viable mass spectra for the exotic Higgs bosons in the model that were consistent with the theoretical constraints of vacuum stability and perturbative unitarity and the experimental constraints of electroweak precision observables, $Zb \bar b$ coupling and Higgs boson signal strengths.  In Fig.~\ref{chipptoH1}, the thick black solid and dashed curves reproduce those in Fig.~1 of Ref.~\cite{Chiang:2015amq}, representing the 1$\sigma$ and 2$\sigma$ contours of a $\chi^2$ fit to the signal strengths of the $W^+W^-$, $ZZ$, $b \bar b$ and $\tau^+\tau^-$ channels of the SM-like Higgs boson measured in LHC Run-I, including the glue-glue fusion (GGF) and vector boson fusion (VBF) production mechanisms.  The signal strength of the diphoton channel is left out because of uncertainties in the mass of and couplings with the charged Higgs bosons.

As already studied and presented in Fig.~3 of Ref.~\cite{Chiang:2015amq}, for each point within the $2\sigma$ contour of Fig.~\ref{chipptoH1} the mass spectrum of the exotic Higgs bosons are still subject to the theoretical constraints of perturbative unitarity and vacuum stability and the experimental constraints of electroweak $S$ parameter and $Z b \bar b$ coupling to have different patterns.  As shown there, after taking the above-mentioned constraints and without other inputs, the possible mass ranges are still fairly large.  In the next section, we will show that the parameter space is significantly reduced if we take $H_1^0$ as the putative 750-GeV resonance and impose the required production rate and total decay width.

Before closing this section, we comment on the possibility of using $H_3^0$ or $H_5^0$ to explain the diphoton excess.  As given in Table~\ref{coupling}, the dominant production of $H_3^0$ ($H_5^0$) at the LHC is the GGF (VBF) mechanism.  Both cross sections are proportional to $v_\Delta^2$, but independent of the mixing angle $\alpha$.  It turns out that both production rates are too small to explain the observed excess even when $v_\Delta$ saturates the electroweak VEV.

\section{Diphoton excess \label{sec:excess}}

The diphoton excess observed by the ATLAS and CMS Collaborations, as quoted in Eq.~(\ref{eq:data}), is averaged to be 
\begin{align}
\sigma(pp\to S \to\gamma\gamma)
&= (5.26 \pm 1.22) \text{ fb}.
\label{eq:avgdata}
\end{align}
ATLAS also reported that the total decay width of the resonance $\Gamma_S \simeq 45$~GeV.  Although these data are not conclusive, it is nevertheless an amusing  exercise to see whether the singlet Higgs boson of the custodial Higgs triplet model can possibly accommodate the data.

\subsection{Production rate of $H_1^0$}

As in the case of $h$, the $H_1^0$ boson can be produced through both GGF and VBF mechanisms at the LHC.  In view of possible enhancement in its coupling with diphotons, we also include the  $\gamma\gamma$ fusion ($\gamma\gamma$F) mechanism in the analysis.  Therefore, the total production cross section of $H_1^0$ is:
\begin{equation}
\sigma(pp\to H_1)_{\text{total}}
= \sigma(pp\to H_1)_{\text{GGF}} + \sigma(pp\to H_1)_{\text{VBF}}  + \sigma(pp\to H_1)_{\gamma\gamma\text{F}} ~.
\end{equation}
In the narrow width approximation, the GGF production cross section of $H_1^0$ is given by
\begin{align}
\sigma(pp\to H_1)_{\text{GGF}}
= \frac{\pi^2}{8} \frac{\Gamma(H_1\to gg)}{M_{H_1}}
\left[\frac{1}{s}\frac{\partial L_{gg}}{\partial\tau} \right] ~,
\nonumber
\end{align}
where $\tau=m_{H_1}^2/s$ and $\sqrt s$ is the center-of-mass colliding energy.  The parton luminosity factor for the $gg$ initial state in the square brackets is taken to be $0.97~(4.4)~ K_g\times10^3$~pb for the 8-TeV (13-TeV) collisions using the MSTW2008 parton distribution functions (PDF's)~\cite{MSTW2008}, where the next-to-leading-order (NLO) K-factor $K_g \simeq 2$~\cite{Kfactor}.  For the VBF part, it is easier to scale it from the 8-TeV simulation for a 750-GeV SM Higgs boson: $\sigma(pp\to H_1)_{\text{VBF}}^{\text{13-TeV}} = 2.496~(\kappa_V[H_1^0])^2 \times \sigma(pp\to h_{SM})_{\text{VBF}}^{\text{8-TeV}}$, where $\sigma(pp\to h_{SM})_{\text{VBF}}^{\text{8-TeV}} = 0.05235$~pb~\cite{CERN}.  For the $\gamma\gamma$F process, we use a formula similar to the GGF process:
\begin{align}
\sigma(pp\to H_1)_{\gamma\gamma\text{F}}
= \frac{\pi^2}{8} \frac{\Gamma(H_1\to \gamma\gamma)}{M_{H_1}}
\left[\frac{64}{s}\frac{\partial L_{\gamma\gamma}}{\partial\tau} \right] ~,
\nonumber
\end{align}
where the parton luminosity factor in the square brackets is taken to be 54 (101)~pb for the 8-TeV (13-TeV) collisions using the NLO NNPDF23 PDF's~\cite{NNPDF}.

Even though $\Gamma(H_1^0 \to W^+W^-) > \Gamma(H_1^0 \to gg)$ for the parameter space of interest to us, the GGF production is still dominant due to the large $gg$ parton luminosity factor.   Since both the decay widths of $H_1^0$ to $gg$ and $WW$ depend only on $\alpha$ and $v_\Delta$, we thus superimpose red contours of the $H_1^0$ production cross section via the GGF and VBF mechanisms at the 13-TeV LHC in Fig.~\ref{chipptoH1}.  It is noted that the red curves are almost symmetric with respect to $\alpha = 0$, reflecting the fact that the GGF production is dominant as its rate is proportional to $\sin^2\alpha$ from the top-quark loop contribution.  As $v_\Delta$ increases, $\cos\beta$ gets larger and so does the $H_1^0$-$W^+$-$W^-$ coupling, particularly when $\alpha \to 0$.  In this case the VBF production also becomes important, resulting in the red elliptical contours in the figure.

The $\gamma\gamma$F part becomes dominant only when $v_\Delta$ approaches zero.  However, due to its complicated dependence, it is impossible to also show its contribution in Fig.~\ref{chipptoH1}.  Nonetheless, we note with caution that the above-quoted values of diphoton parton luminosity factor are the central values and suffer from large uncertainties (more than $\sim 30\%$ for the photon PDF when the momentum fraction is greater than $0.001$).

\subsection{Imposing constraints from 8-TeV data}

Before fitting to the required diphoton rate at 13 TeV, we first consider the existing constraints on the production of the 750-GeV $H_1^0$ in the following channels from the 8-TeV collision data:
\begin{align}
\begin{split}
\sigma(pp\to H_1^0 \to\gamma\gamma)
&< 1.5~{\rm fb}~\mbox{\cite{diphoton1,diphoton2}} ~,
\\
\sigma(pp\to H_1^0 \to WW)
&< 40~{\rm fb}~\mbox{\cite{WW1,WW2}} ~,
\\
\sigma(pp\to H_1^0 \to ZZ)
&< 12~{\rm fb}~\mbox{\cite{ZZ}} ~,
\\
\sigma(pp\to H_1^0 \to Z\gamma)
&< 11~{\rm fb}~\mbox{\cite{Zphoton}} ~,
\\
\sigma(pp\to H_1^0 \to jj)
&< 2.5~{\rm fb}~\mbox{\cite{jj1,jj2}} ~.
\end{split}
\label{8TeV}
\end{align}
The jets $jj$ in the last inequality above include pairs of gluons and quarks except for $t \bar t$.  We find that the bounds in Eq.~(\ref{8TeV}) hardly constrain our parameter space, except that $\sigma(pp\to H_1^0\to\gamma\gamma)$ can exclude a very small $v_\Delta$ region, as will be shown below.

\begin{figure}[ht]
\centering
\includegraphics[scale=0.37]{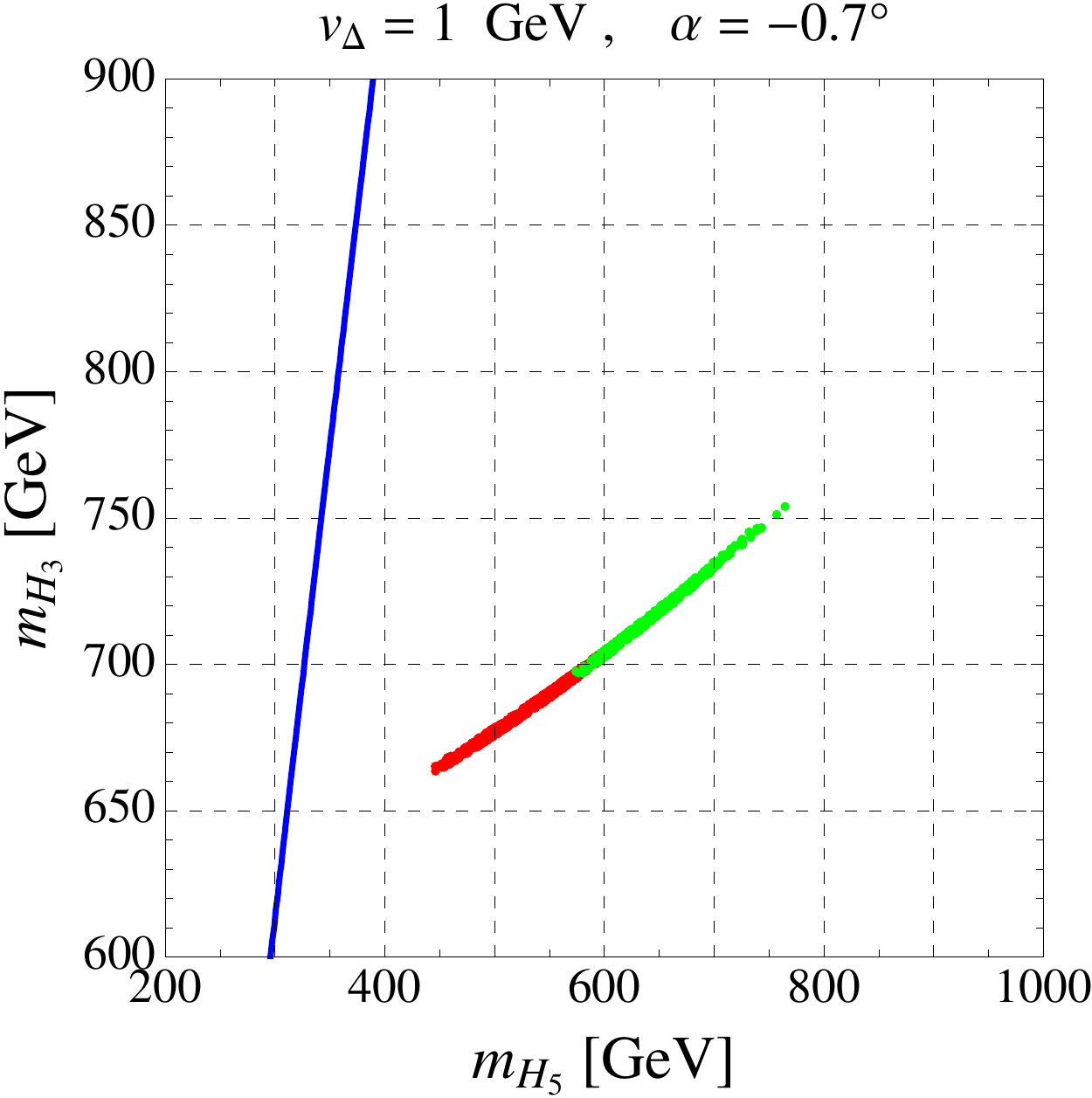}
\includegraphics[scale=0.37]{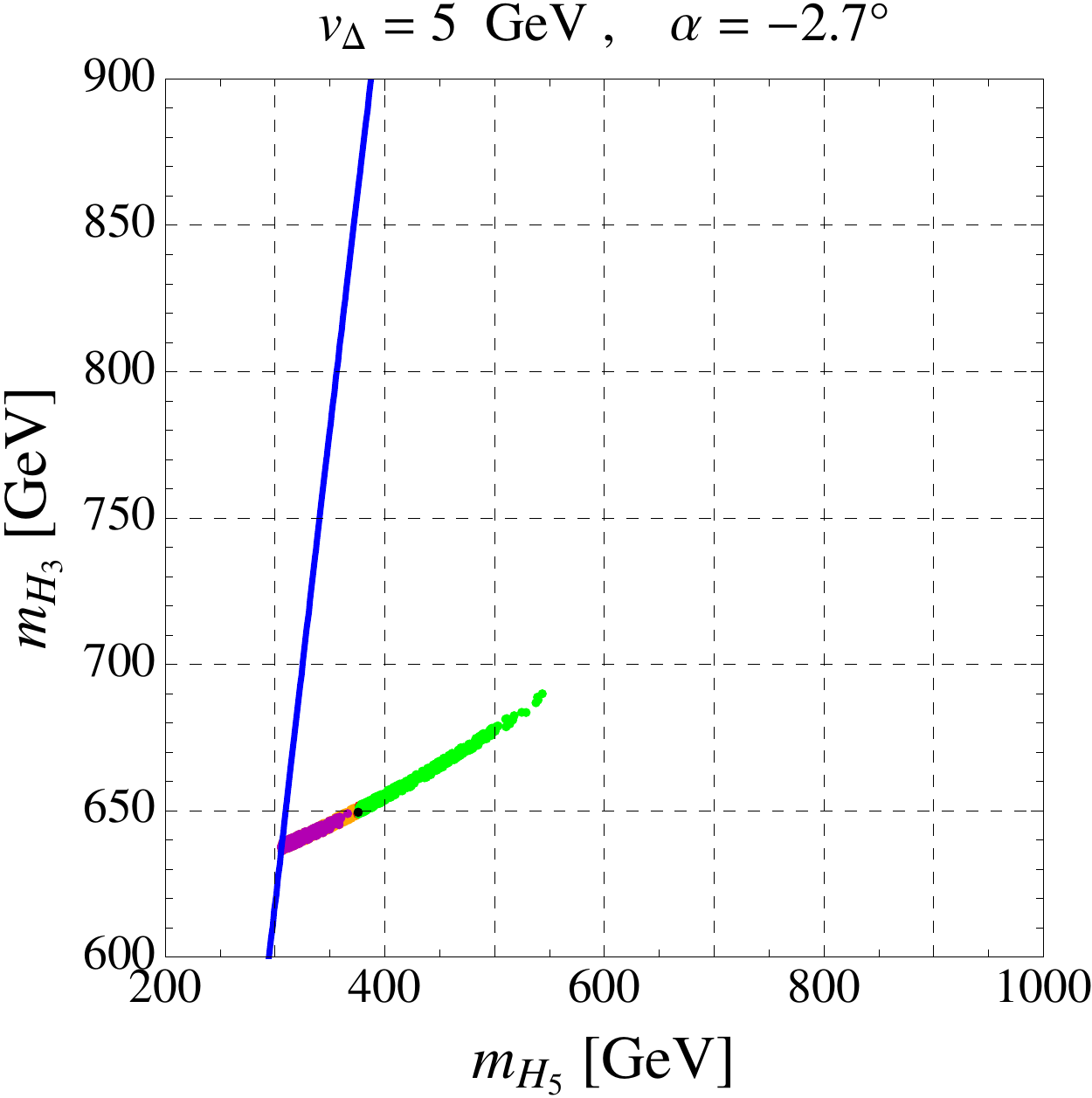}
\includegraphics[scale=0.37]{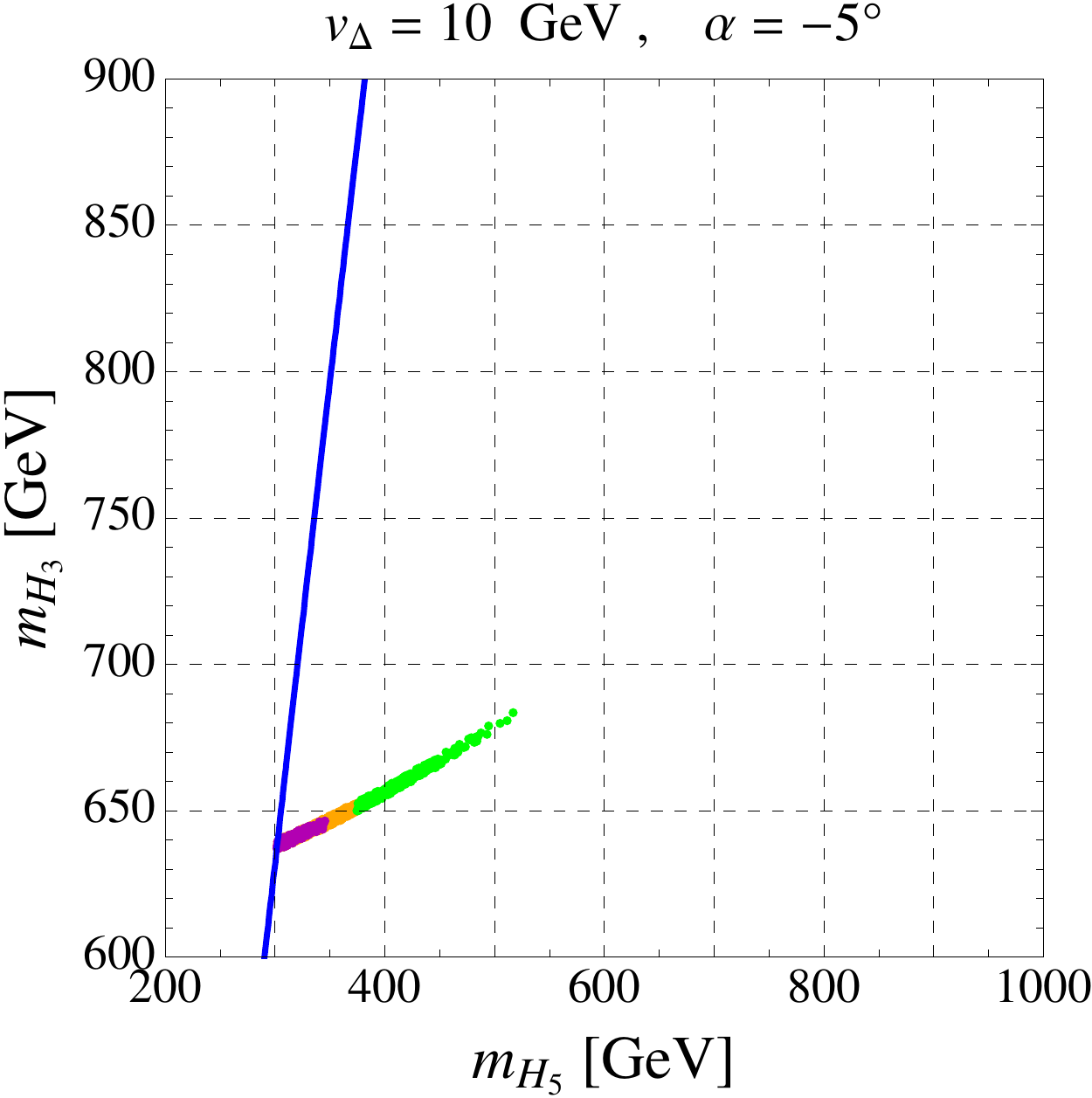}
\includegraphics[scale=0.37]{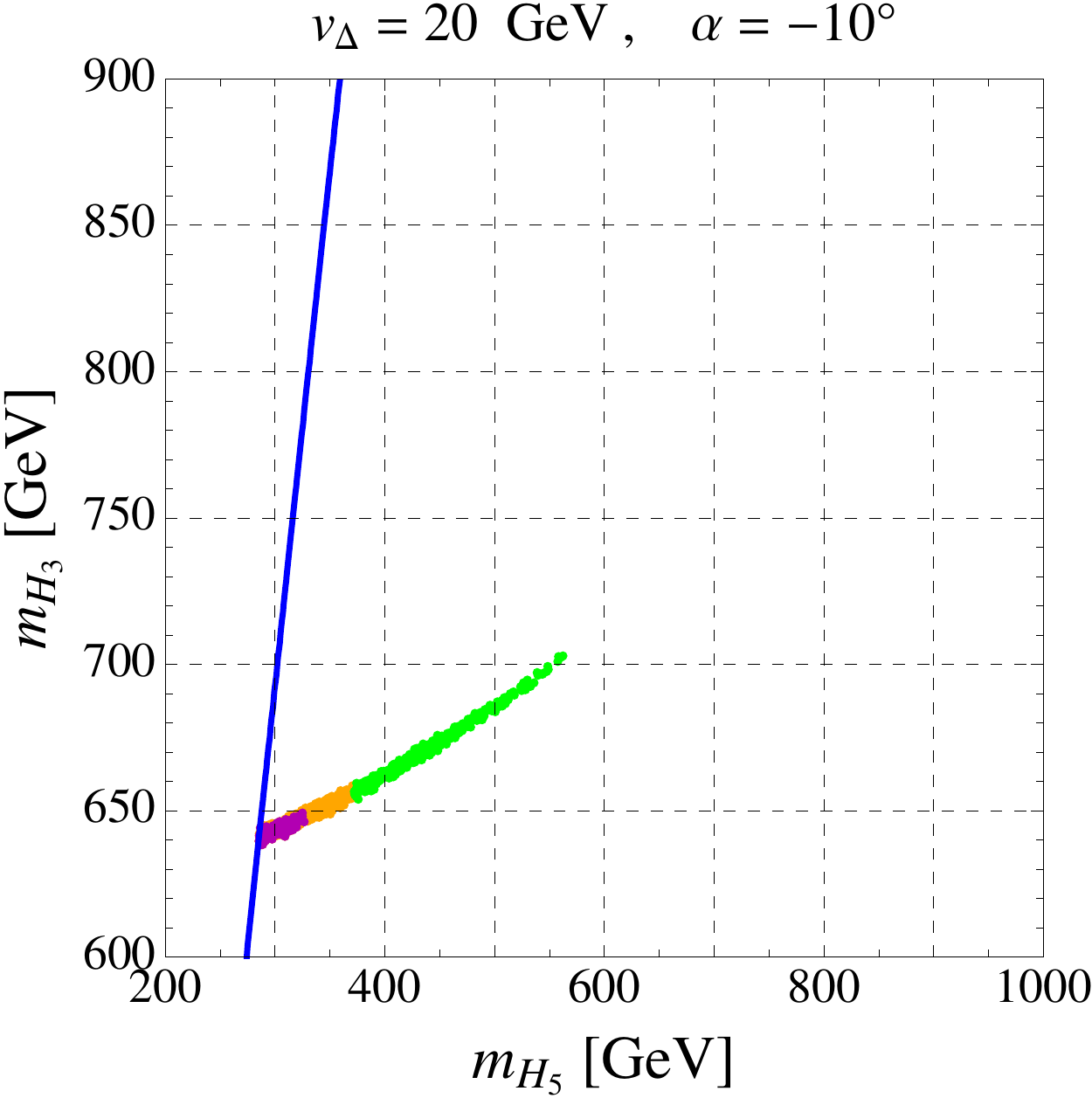}
\vspace{2mm}
\includegraphics[scale=0.37]{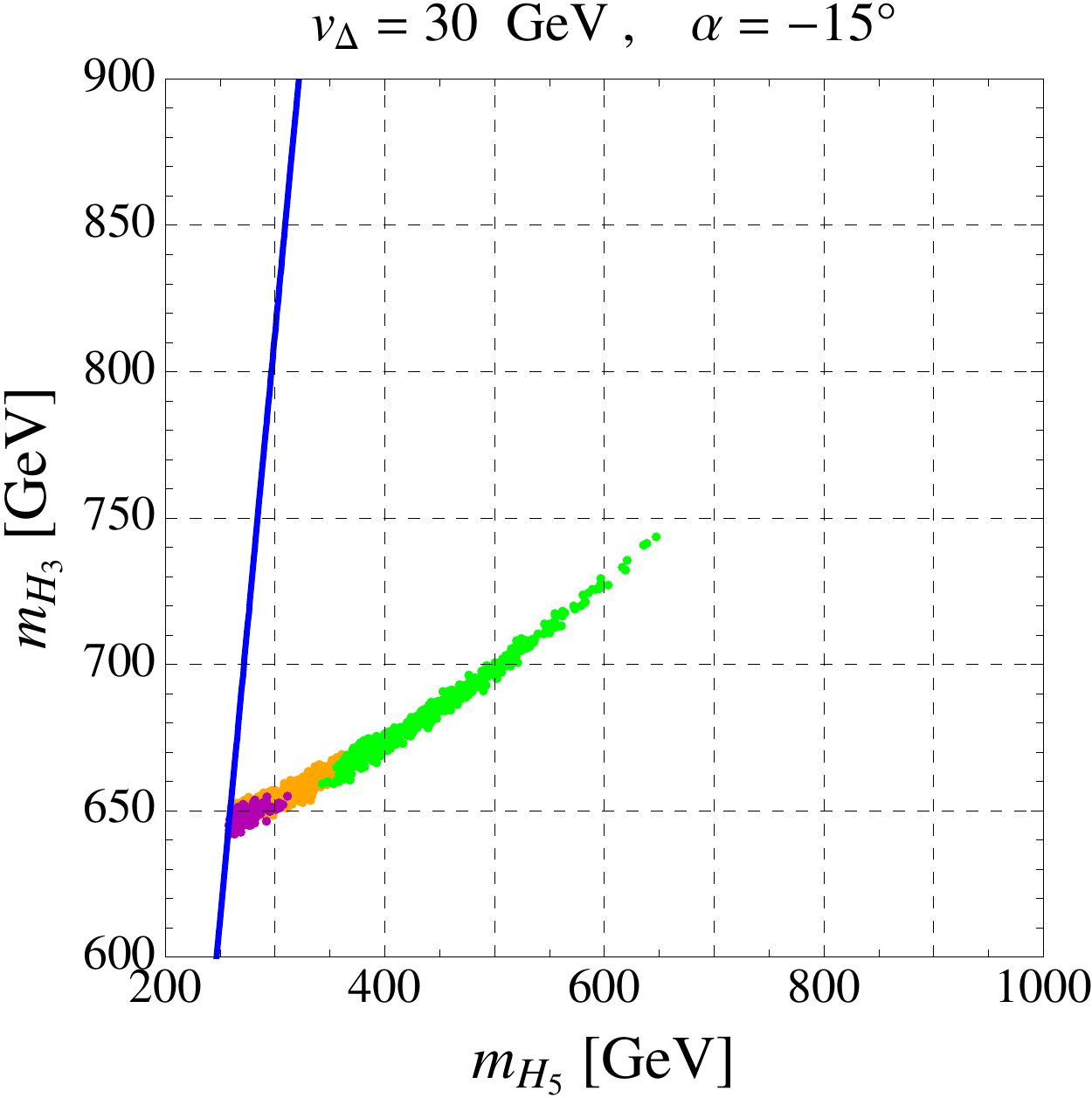}
\includegraphics[scale=0.37]{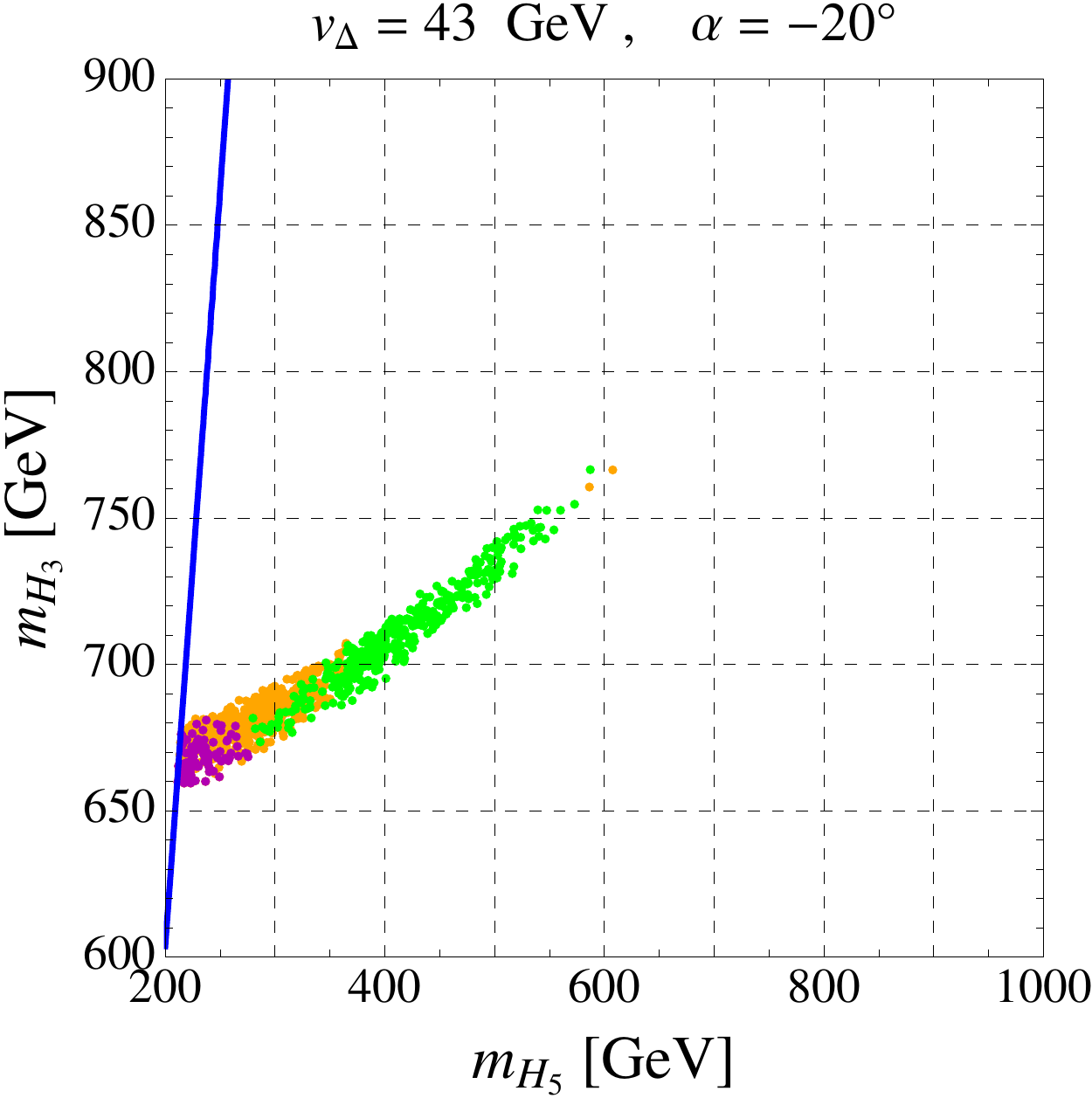}
\caption{Representative scatter plots of viable mass spectra.  Colored points are those passing the constraints analyzed in Ref.~\cite{Chiang:2015amq} and $m_{H_1} = 750$~GeV.  The purple dots are ruled out by $\mu_{\gamma\gamma}$ from LHC Run-I at 95\% CL.  The orange dots are excluded by $\Gamma_{H_1^0} < 50$~GeV.  The red dots are excluded by the 8-TeV data of $\sigma(pp\to H_1^0\to\gamma\gamma)$.  The parameter space to the right of the blue lines are allowed by the $S$ parameter. The green dots give the viable mass spectra satisfying all the bounds and constraints explained in the main text, with only the black dots being able to explain the 750-GeV diphoton resonance within $2\sigma$.}
\label{NLO}
\end{figure}

Fig.~\ref{NLO} gives for the 6 blue benchmark points in Fig.~\ref{chipptoH1} and $m_{H_1} = 750$~GeV the scatter plots of $m_{H_3}$ and $m_{H_5}$ allowed by the constraints already analyzed in Ref.~\cite{Chiang:2015amq}.  Here we note that $m_{H_3}$ and $m_{H_5}$ present a correlation in the colored dots.  For example, in the decoupling limit of $v_\Delta \to 0$ and $\alpha \to 0$, we have an approximate relation:
\begin{align}
m_{H_1}^2
= \frac12 (1 - \alpha^2) \left( 3m_{H_3}^2-m_{H_5}^2 \right) + \alpha^2 m_h^2 ~.
\end{align}
In order to justify the narrow width approximation and to be consistent with current data, we further require that the total width of $H_1^0$ is less than 50~GeV.  This rules out the orange dots in the plots.  Most of the orange dots concentrate in the low $m_{H_5}$ region because the $H_1^0$ decays to a pair of $H_5$ of different charges open up, increasing the total width.  There are also a few orange dots in the upper $m_{H_5}$ region in the plot for $(v_\Delta,\alpha)=(43, -20^\circ)$.  This is because the $H_1\to hh$ decay has a larger width.  The purple dots are ruled out by the constraint of $\mu_{\gamma\gamma}$, the signal strength of $pp \to h \to \gamma\gamma$, from LHC Run-I at 95\% confidence level (CL).  The excluded purple dots form a subset of the orange dots, as the charged Higgs bosons are sufficiently light to enhance the decay.  The red dots, appearing only in the cases of $v_\Delta \alt 1$~GeV, mark those excluded by the 8-TeV data of $\sigma(pp\to H_1^0\to\gamma\gamma)$.  In the end, only the green dots pass all the above-mentioned requirements, and only the black dots can possibly explain the 750-GeV diphoton excess at 13 TeV.

From such a parameter scan, we observe that there exists lower bounds on $m_{H_{3,5}}$, which are derived from a compilation of different constraints: unitarity, stability, electroweak $S$ parameter, and $\sigma(pp\to H_1^0\to\gamma\gamma)$ at 8 TeV.  For a given $v_\Delta$, the allowed mass spectra become lower for larger $\alpha$ and an upper bound on $\alpha$ exists, beyond which some of the above-mentioned constraints are violated.  We therefore identify that points in the linear region lying along the 6 benchmark points in Fig.~\ref{chipptoH1} deserve closer investigations.  We note in passing that the partial width of $H_1^0 \to hh$ is proportional to $|\alpha|^2$.  Therefore, $|\alpha|$ cannot be too large; otherwise the total width of $H_1^0$ exceeds the limit of $50$~GeV imposed in this work.

\begin{table}
\begin{center}
\begin{tabular}{cccccccccc} \hline \hline
$(v_\Delta / {\rm GeV}, \alpha)$
 & $(43, -20^\circ)$  & $(30, -15^\circ)$ 
& $(20, -10^\circ)$ & $(10, -5^\circ)$  &$(5, -2.7^\circ)$ &$(1, -0.7^\circ)$ \\
\hline
$\sigma_{\rm total}$ [ab]  & 
$0-1$          & $0-12$       & $0-33$     & $0-186$ &$0-2848$ &$0-2822$\\ 
$\sigma_{\rm GGF}$ [ab] & 
 $0-1$         &$0-10$        &$0-26$       & $0-74$  &$0-130$   &$0-73$\\
$\sigma_{\rm VBF}$ [ab]  & 
 $0-10^{-1}$&$0-2$          &$0-5$        &$0-15$    &$0-19$   &$0-3$\\
$\sigma_{\gamma\gamma{\text F}}$ [ab] & 
 $0-10^{-4}$ &$0-10^{-1}$&$0-1$         &$0-97$  &$0-2698$ &$0-2746$\\
\hline
$\Gamma_{H_1}$ [GeV]  &
$46-50$       &  $20-50$   & $9-46$      & $2-23$   &$1-4$    &$0-0.1$\\
${\cal B}(WW)$ &
 $55-59$      &$24-59$      & $11-59$    & $6-58$  &$8-55$  &$10-39$\\
${\cal B}(ZZ)$ &
$27-29$       & $12-29$     & $6-29$      & $3-28$  &$4-27$  &$5-19$\\
${\cal B}(t\bar t)$&
$9-10$         & $5-11$       & $2-10$       & $1-10$  &$2-13$ &$8-30$\\
${\cal B}(hh)$&
 $2-8$          &$0-8$           & $0-6$        & $0-5$    &$0-2$ &$4-13$\\
${\cal B}(\gamma\gamma)$& 
$0-10^{-3}$  & $0-10^{-2}$ &$0-10^{-1}$ & $0-1$    &$0-7$  &$0-57$\\
${\cal B}(Z\gamma)$            &
$0-10^{-3}$  &$0-10^{-3}$   &$0-10^{-2}$&$0-10^{-1}$&$0-1$  &$0-10$\\ 
${\cal B}(gg)$                        & 
$\sim10^{-2}$&$\sim10^{-2}$&$0-10^{-2}$&$0-10^{-2}$&$\sim10^{-2}$ &$0-10^{-1}$    \\ 
${\cal B}(H_3^+W^-)$            & 
$0-10^{-2}$  &$0-1$              & $0-6$        & $0-23$   &$0-35$     &$0-4$\\
${\cal B}(H_5^{++}H_5^{--})$& 
$0-3$            &$0-24$           & $0-31$      & $0-32$   &$0$   &$0$\\ 
\hline\hline
\end{tabular}
\caption{Total, GGF, VBF, and $\gamma\gamma$F production cross sections of $pp\to H_1\to\gamma\gamma$ at the 13-TeV LHC in units of ab, total decay width of $H_1$ and the branching ratios of various $H_1$ decays in units of \%, calculated for several sets of $(v_\Delta, \alpha)$.  The sum of all the $H_5$ pair modes is 2.5 times that of ${\cal B}(H_5^{++}H_5^{--})$.}
\label{widthBR}
\end{center}
\end{table}

In the following, we focus on the parameter regions represented by the green dots in Fig.~\ref{NLO}.  Since $H_1^0$ is a CP-even Higgs boson, its possible decay channels include the $WW$, $ZZ$, $gg$, $q\bar{q}$, $\ell\bar\ell$, $\gamma\gamma$, $Z\gamma$, $H_3^\pm W^\mp$, $hh$, $H_5H_5$ and $H_3H_3$ final states.  The branching ratios of important channels are given in Table.~\ref{widthBR}.  In most of the allowed parameter space, the $H_1^0 \to H_3H_3$ and $H_5H_5$ channels of different charges are kinematically forbidden.  This is largely because these decay modes will significantly contribute to the total width so that $\Gamma_{H_1^0} > 50$~GeV.  The dominant decays of $H_1^0$ are thus mostly the $WW$ and $ZZ$ modes.  It is a prediction of the model that the $WW$ channel is twice stronger than the $ZZ$ channel.  For $v_\Delta \agt 10$~GeV, $\sigma(pp \to H_1^0)$ ranges from a few up to about $100$~ab.  To get a diphoton cross section of a few fb, one expects that ${\cal B}(H_1^0 \to \gamma\gamma)$ has to reach at least the percent level, which, according to Table~\ref{widthBR}, happens only when $v_\Delta < 10$~GeV.

\subsection{Small $v_\Delta$ region}

We now scrutinize the small $v_\Delta$ region where the diphoton coupling of $H_1^0$ can become sufficiently strong to enhance the 750-GeV diphoton signal.  Such an enhancement in the coupling mainly comes from the $H_5^{\pm\pm}$ loop, particularly when $m_{H_5}$ is close to $m_{H_1}/2$.  To maximize this effect, it is desirable for $(v_\Delta, \alpha)$ to approach the origin along the linear region indicated by the blue dots in Fig.~\ref{chipptoH1}, until the 8-TeV data of $\sigma(pp\to H_1^0\to\gamma\gamma)$ pushes the viable mass spectra upward for $v_\Delta \alt 1$~GeV.  Generally speaking, as $v_\Delta$ gets smaller, ${\cal B}(H_1^0\to\gamma\gamma)$ becomes larger, possibly leading to a sufficiently large diphoton production rate for the observed data.  However, as noted before, the production rate for small $v_\Delta$, as those given in the last two columns of Table~\ref{widthBR}, suffers from large uncertainties in the photon PDF of the proton.  For example, it can be easily invalidated by reducing the photon parton luminosity factor by $50\%$.  In addition, we provide an argument below to refute the apparent possibility of explaining the diphoton excess in this region.

To reach the required $\Gamma(H_1^0\to\gamma\gamma)$, the dimensionful $H_1H_5H_5$ triple Higgs coupling $g_{H_1H_5H_5}$ is generally of about a few tens of TeV, implying that the dimensionless coupling $c_{\gamma\gamma}$ appearing in the effective diphoton interaction with $H_1^0$
\begin{align}
{\cal L}_{\rm eff} = \frac{e^2}{4v_{\Phi}} c_{\gamma\gamma} H_1 F_{\mu\nu} F^{\mu\nu}
\end{align}
is about ${\cal O}(\bf\rm 0.4-0.7)$.  Though such values for $c_{\gamma\gamma}$ seem to support perturbative calculations, we find that $g_{H_1H_5H_5}$ gets a strong constraint when one checks the perturbation series in the quartic $H_1$ coupling.  In the same parameter region, the tree-level quartic $H_1$ coupling $\lambda_{H_1H_1H_1H_1} \sim {\cal O}(0.02-0.5)$.  Yet the one-loop correction to the vertex due to the box diagram mediated by the $H_5$ bosons
\begin{align}
\delta\lambda_{H_1H_1H_1H_1} \sim \frac{1}{16 \pi^2} \frac{g^4_{H_1H_5H_5}}{8 m_{H_5}^4}
\sim {\cal O}(10^3-10^4) ~.
\end{align}
Apparently, this signifies the breakdown of perturbation.

\section{Conclusions \label{sec:summary}}

Even though the data are yet inconclusive, the recent observation of diphoton excess around the mass of 750~GeV in LHC Run-II by both the ATLAS and CMS Collaborations inspires us to examine whether the singlet Higgs boson of the custodial Higgs triplet model can serve as a good candidate.  Such a model is motivated to give Majorana mass to neutrinos through $SU(2)_L$ Higgs triplet fields while preserving the custodial symmetry.  Under the $SU(2)_L$ symmetry, the exotic Higgs bosons originating from the triplet fields are decomposed into a singlet, a triplet, and a quintet, with the former two being able to mix with the corresponding representations from the standard model Higgs doublet field.  When the triplet fields acquires an ${\cal O}(1)$~GeV or larger vacuum expectation value as induced by the breakdown of electroweak symmetry, the exotic Higgs bosons exhibit novel collider phenomena.

Based on an earlier study of comprehensive parameter scan for viable mass spectra, we show that a 750-GeV $H_1^0$ falls well within the allowed parameter space.  Imposing the constraints of 8-TeV search data for the assumed mass, we find several benchmark points of $(v_\Delta,\alpha)$ lying along a line passing through the origin to be of interest, where $v_\Delta$ denotes the triplet vacuum expectation value and $\alpha$ is the mixing angle between the heavy Higgs singlet and the 125-GeV Higgs boson. 
Moreover, a definite mass hierarchy, $m_{H_1} = 750~{\rm GeV} > m_{H_3} > m_{H_5}$, emerges among the exotic Higgs bosons in this region of parameter space.  Such information enables us to make more definite predictions about how the exotic Higgs bosons decay.

We have worked out the possible ranges of production rates, total decay width and branching ratios of the $H_1^0$ boson for the benchmark points.  In particular, we find that the maximum diphoton production through $H_1^0$ is ${\cal O}(1-100)$~ab, 1 to 3 orders of magnitude off the observed 750-GeV diphoton excess at LHC Run-II, except for the small $v_\Delta$ region.  In such a region, however, we find that the $H_1H_5H_5$ coupling becomes so large that the quartic $H_1^0$ coupling does not grant good perturbation, not to mention the large uncertainties in the photon parton distribution function of the proton.  Therefore, we conclude that the heavy Higgs singlet boson in the custodial Higgs triplet model cannot explain the observed diphoton excess at 750~GeV.

\acknowledgments

The authors would like to thank Eibun Senaha for useful discussions.  This work is supported in part by the Ministry of Science and Technology of Taiwan under Grant No.~MOST104-2628-M-008-004-MY4.

\end{document}